\documentclass{elsarticle}
\usepackage{graphicx}
\usepackage[usenames]{color}
\begin{document}
\begin{frontmatter}

\title{The interaction of Fe thin layers between MgO(100)-MgO and MgO(100)-Ag surfaces.}
\author[Budapest]{I. D\'{e}zsi\corref{cor1}\fnref{*}}
\ead{dezsi@kfki.rmki.hu}
\author[Budapest]{Cs. Fetzer}
\author[Budapest]{F. Tanczik\'{o}}
\author[Krakow]{J. Korecki}
\author[Otsu]{A. Nakanishi}
\author[Otsu]{T. Kobayashi}
\address[Budapest]{KFKI Research Institute for Particle and Nuclear Physics H-1525
Budapest 114 P.O. Box 49, Hungary}
\address[Krakow]{Institute of Catalysis and Surface Chemistry
Polish Academy of Sciences 30-239 Krakow, Poland}
\address[Otsu]{ Shiga University of Medical Science
Seta, Otsu, Physics Department, 520-2192 Shiga, Japan}
\cortext[cor1]{ Corresponding author}
\begin{abstract}
The atomic interaction and magnetic properties
of ultrathin Fe films grown on cleaved and polished MgO(100) surfaces
were studied by conversion electron M\"{o}ssbauer spectroscopy (CEMS) in broad temperature range.
Fe with different  layer thickness was deposited on MgO substrates.
The layers were formed on polished and cleaved substrate surfaces at RT. The analysis of the spectra showed no Fe-O$^{2-}$ interaction in MgO/Fe interface. Iron layers showed different magnetic anisotropy depending on their thickness.
\end{abstract}
\end{frontmatter}
\section{Introduction}

Magnetic tunnel junctions are the key effect in the development of
magnetoelectronic devices. They operate using as spin-dependent
tunnel magnetoresistance  effect. The Fe/MgO/Fe(001) system is
currently under scope since theoretical calculations predicted a
very high tunneling magnetic resistance\cite{Butler}.  Numerous experimental and theoretical works
showed that the key role of the nature of tunneling processes are
the electronic structure at the metal - insulator interface. Some
authors observed FeO formation on the surface
\cite{Tusche,Meyerheim1,Meyerheim2,Zhang,Wang}, others observed
and calculated no or very weak interaction between Fe and O$^{2-}$
ions\cite{Li,Luches,Sicot1,Sicot2,Plucinski,Zajac} depending on the preparation mode.
Therefore, the determination of the interaction between Fe and the surrounding
layers is, an important and open question.
M\"{o}ssbauer spectroscopy using $^{57}$Fe  offers
the real suitable technique to determine the atomic interactions
at the interfaces  \cite{Shinjo1,Shinjo2} and the
results may explain the reason of the discrepancies existing in
earlier studies.   Fe film can be deposited on the substrate and covered by  different layers. We have chosen  MgO(100) substrate and covered the deposited $^{57}$Fe by MgO or Ag layers.
Fe and Ag is immiscible and the interaction between the layers earlier has been studied \cite{Tyson, Shurer,Dezsi}.  Our main aim was to
determine the interaction of Fe on the MgO/Fe/MgO  interfaces on polished and cleaved MgO(100) substrates. Also,  the deposited iron was also covered with Ag to compare  the difference of MgO and Ag cover.  The measurements were made down to 15 K.
\section{Experimental Details}
The $^{57}$Fe layers were deposited on the surface of polished and cleaved
(ex situ) MgO(100) single crystals cleaned properly. The
substrates were UHV annealed prior to the depositions in the
pretreatment chamber of the MBE (MECA 2000) system at 990 K for 30
min. The base pressure in the chambers was 1x10$^{-10}$ Torr and
increased to 2x10$ ^{-9}$ Torr during the depositions. Series of
samples containing  3 to 10 ML Fe.
$^{57}$Fe were grown at the rate of 0.025 ML/s.  Multilayer $^{57}$Fe samples
were deposited first on the MgO(100) substrates and covered by MgO.
Other samples were  capped with
10 nm thick Ag and 5 nm thick Si layers to prevent any oxidation
of the Fe surface in ex situ measurements.
 {$^{57}$Fe  was evaporated by using Knudsen-cells with
BeO crucibles.\newline The M\"{o}ssbauer measurements
were carried out by using a conventional constant
acceleration-type spectrometer. For the detection of the
conversion electrons a low-background proportional counter filled
with H$_2$ was used. The spectra were measured by 50
mCi $^{57}$Co(Rh) single line source. For the analysis of the spectra, a
least-squares fitting program was used. Also, using this program,
spectra with histogram distributions of parameter values could be
fitted. The spectra with internal field
distributions were fitted by 35 subspectra in the range of 20 and 45 T.
The linear correlation of the magnetic field with isomer shift was included.
The linewidth was fixed to 0.25 mm/s. The isomer shift values are
given relative to that of $\alpha$-Fe at room temperature.

\section{Results and Discussions}

The M\"{o}ssbauer spectra of the MgO/Fe/MgO samples measured at 15 K are are displayed in Fig. 1.

\begin{figure}[h]
\centering
\includegraphics*[width=5cm]{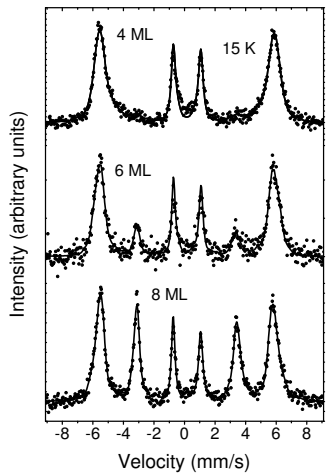}
\caption{M\"{o}ssbauer spectra of  $^{57}$Fe layers between MgO (polished) and deposited MgO measured at 15 K.}
\label{Fig1}

\end{figure}

The spectra are magnetically split and show  thickness dependence.  The samples  were deposited on cleaved substrates because  those samples deposited on polished substrate were rather  complex even at 15 K because of the short magnetic relaxation time as shown for the 4 ML thick Fe sample in Fig. 2.
\begin{figure}[h]
\centering
\includegraphics*[width=5cm]{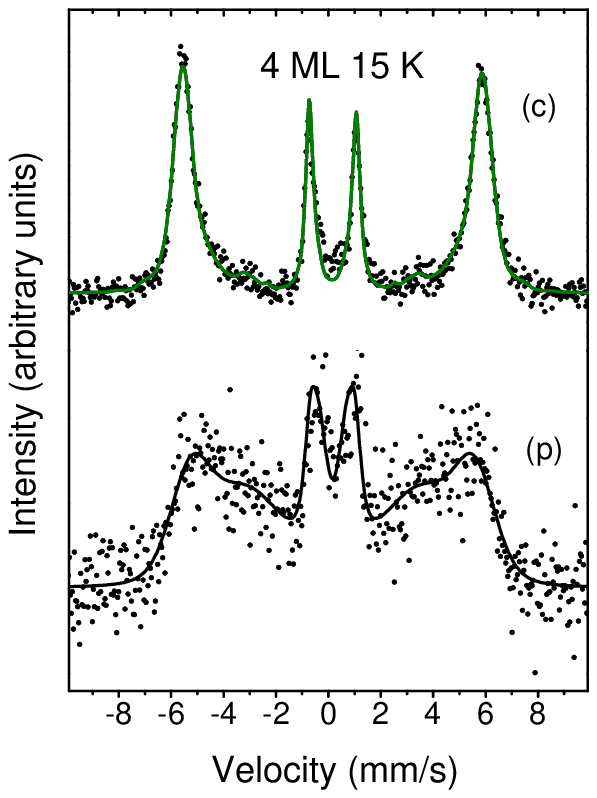}
\caption{M\"{o}ssbauer spectra of 4 ml thick $^{57}$Fe layer on cleaved  (c) and polished (p) substrate.}
\label{Fig2}

\end{figure}

Nevertheless, the 50 nm thick sample deposited at 150 K showed spectrum at RT indicating epitaxial layer formation. The temperature dependence of the spectrum of 8 ML thick Fe sample is shown in Fig. 3.
\begin{figure}[h]
\centering
\includegraphics*[width=5cm]{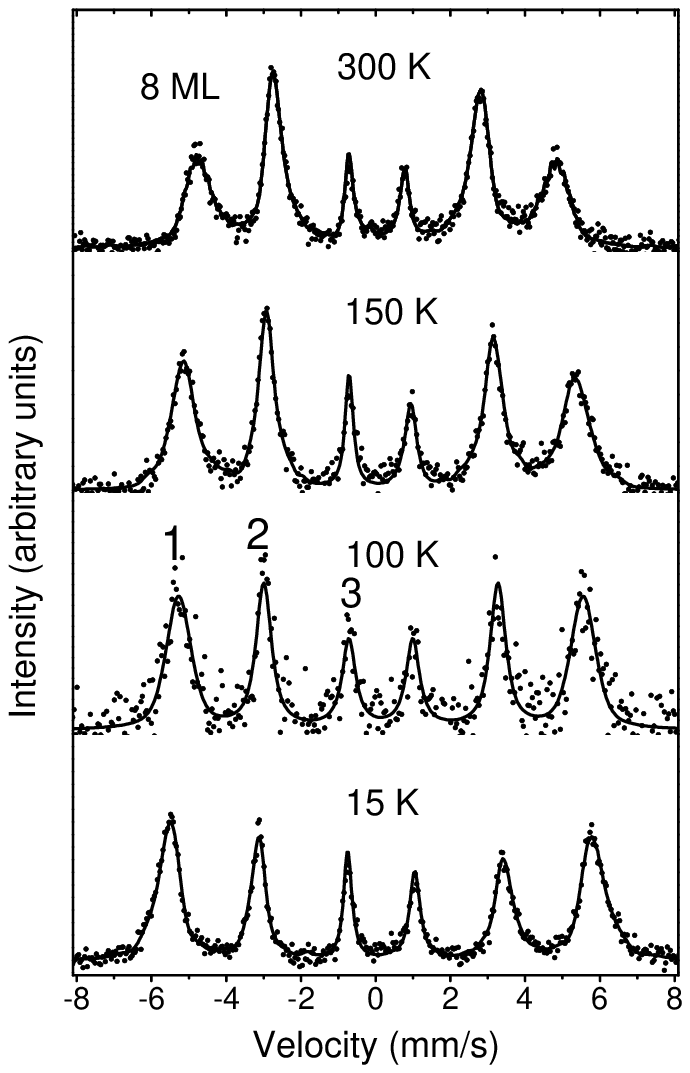}
\caption{M\"{o}ssbauer spectra of 8 ml $^{57}$Fe layer between MgO and MgO at different temperatures.}
\label{Fig3}

\end{figure}

The temperature dependence of the second and third lines ratio is displayed if Fig. 4.

\begin{figure}
\centering
\includegraphics*[width=5cm]{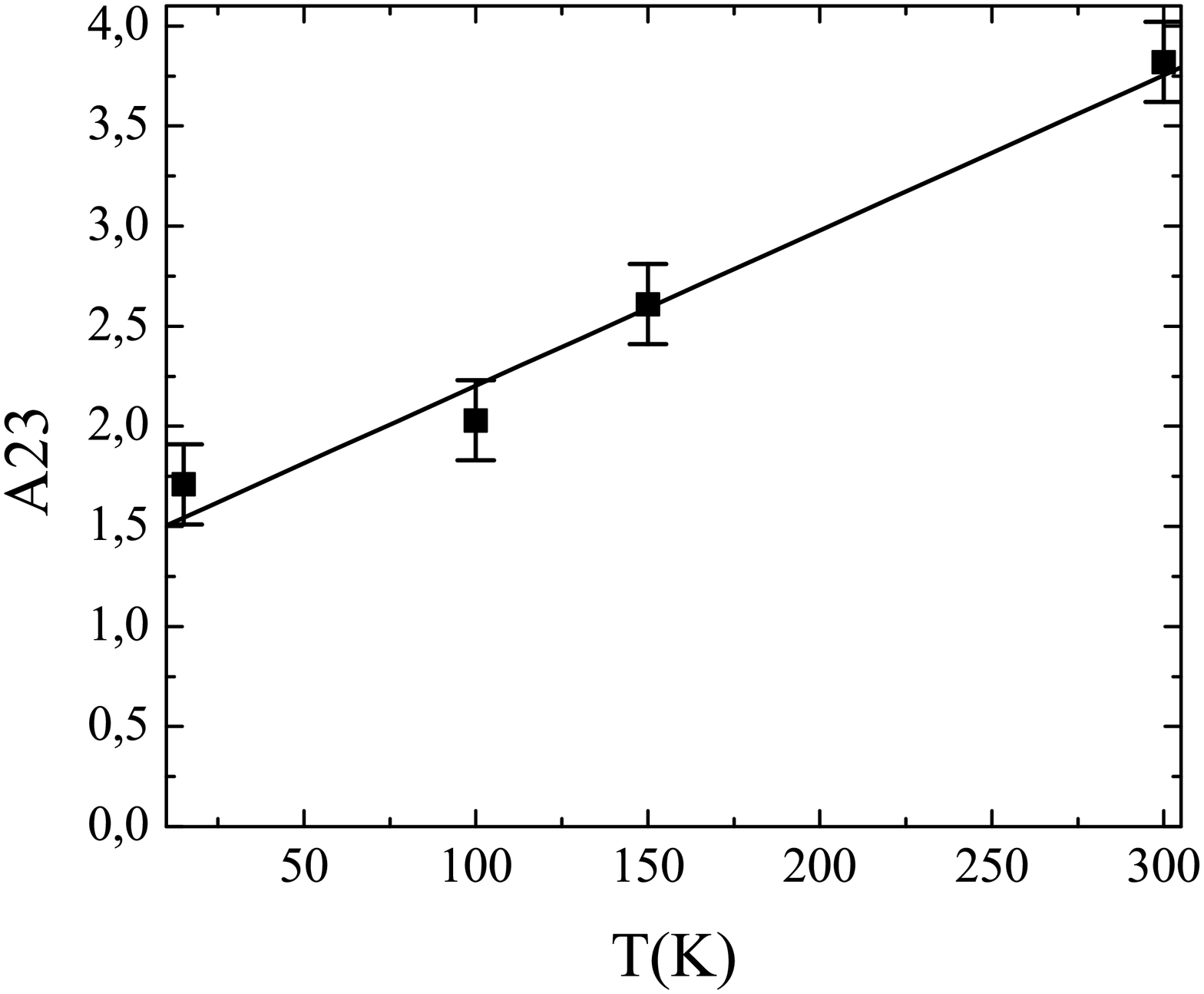}
\caption{Temperature dependence of the intensity ratio of lines 2 and 3 in the  spectrum of 8 ml $^{57}$Fe layer.}
\label{Fig4}

\end{figure}

The spectra of 10 ML thick Fe layer measured at RT and at 15 K are shown in Fig. 5.

\begin{figure}
\centering
\includegraphics*[width=5cm]{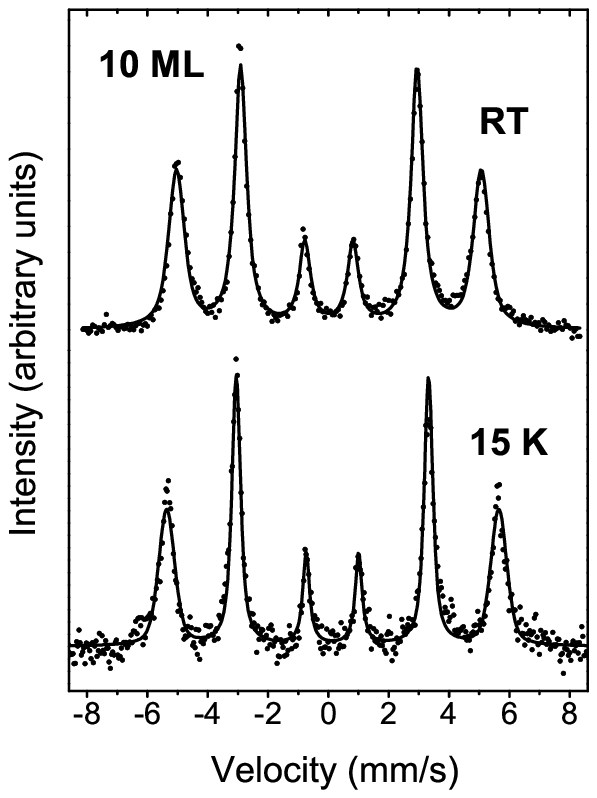}
\caption{M\"{o}ssbauer spectra of 10 ml $^{57}$Fe layers between MgO and MgO.}
\label{Fig5}

\end{figure}

Although, the width of the spectral lines indicate some distribution of the internal magnetic field  only one main component is present for the MgO/Fe/MgO samples. The parameters are compiled in Table I. This is not the case for MgO/Fe/Ag samples shown in Fig. 6. where another component appears with broad line width. All these spectra were fitted by magnetic split components.\newline

\begin{figure}
\centering
\includegraphics*[width=5cm]{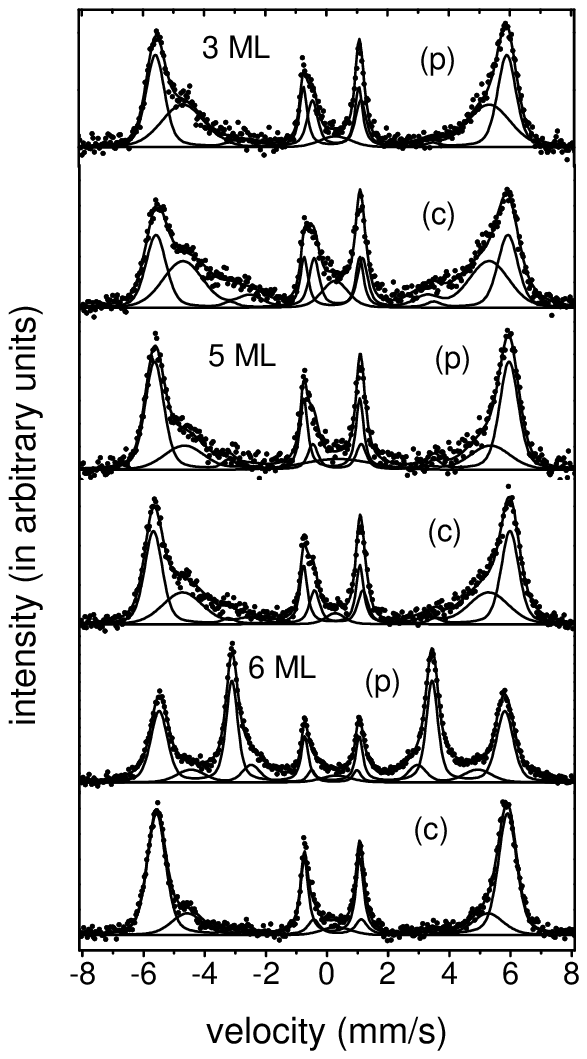}
\caption{M\"{o}ssbauer spectra of  $^{57}$Fe layers between MgO and Ag at different temperatures. c: cleaved, p: polished substrate.}
\label{Fig6}

\end{figure}

\begin{table}
\small
\caption{\label{3}M\"{o}ssbauer parameters of Fe layers on MgO.  Isomer shift ($\delta$) is given in mm/s. The average values values denoted by $< >$. The average hyperfine
magnetic field $<$B$_{hf}$$>$ and the standard deviation (STD) values are
given in Tesla.  A$_{23}$ stands for the relative intensity of the second an third lines. (RI) values for different components are
in percent of the total intensity. (c), (p) denotes cleaved and polished  substrates, respectively. Temperature (T) is given in Kelvin}
\label{tbl}
\begin{tabular}{cccccc}
sample &T& ${<\delta>}$ &{$<$B$_{hf}$$>$ }& STD & A$_{23}$ \\
\hline
MgO(c)/4 ML Fe/MgO&15&0.17&35.1&2.4&0.16(7)\\
MgO(p)/4 ML Fe/MgO&15&0.16&26.9&9.2&0.13(7) \\
MgO(c)/6 ML Fe/MgO&15&0.16&35.4&3.4&0.53(6) \\
MgO(c)/8 ML Fe/MgO&300&0.03&29.6&2.9&3.82(8)\\
&150&0.11&32.5&3.5&2.61(8)\\
&100&0.14&33.6&3.4&1.95(8) \\
&15&0.15&35.2&3.3&1.71(9)\\
MgO(c)/10 ML Fe/MgO&15&0.14&34.2&2.8&2.78(8)\\
\hline
\end{tabular}
\end{table}

Different line intensity ratios were obtained indicating different spin directions depending on the iron thickness. The relative intensity of the Zeeman sextet, 3:R:1:1:R:3 gives information on the angle $\theta$ between the spin orientation and the incident $\gamma$ directions:
\begin{equation}
 \theta  = \arccos \sqrt {{\textstyle{{4 -
R} \over {4 + R}}}}.
\end{equation}

R and   $\theta$ values can be in the range  of 0-4 and 0-90$^\circ$ relative to the crystal normal, respectively. For 3, 5 and 6 ML thick iron layers the spin orientation is perpendicular to the surface.  Nevertheless, second and fifth resonance lines appear for 6 ML thick iron sample deposited in polished substrate. For the sample of 8 ML thick Fe the spin direction is different at RT indicating parallel with surface. At lower temperature, the relative line intensities are changing depending on the temperature. But even at 15 K, the relative intensity has value corresponding to spin directions in between perpendicular and parallel position relative to the surface (Fig. 4.).\newline
The spectra of MgO/Fe/Ag layers depending on $^{57}$Fe thickness are presented in Fig. 6. The fitted hyperfine parameters are displayed in Table II.

\begin{table*}
\centering
\small
\caption{M\"{o}ssbauer parameters of Fe layers between MgO and Ag measured at  15 K. Average isomer shift $<$($\delta$)$>$ is given in mm/s. The average hyperfine
magnetic field $<$B$_{hf}$$>$ and the standard deviation (STD) values are
given in Tesla. Relative intensity (RI) values for different components are
in percent of the total intensity. (c), (p) denotes cleaved and polished  substrates, respectively.}
\begin{tabular*}{\textwidth}{@{\extracolsep{\fill}}lllllll}
sample &components & $<$($\delta$)$>$ &$<$B$_{hf}>$&STD & A$_{23}$ &RI\\

\hline
MgO(p)/3 ML Fe/Ag&A & 0.16 & 35.7& 1.7(1) &0.14(8) & 46.8(9)\\
                 &  B & 0.33 & 30.9 & 4.4(2) &0.27(8) & 50.2(9)\\
                 &  C & 0.32 &  &  & & 3.0(9)\\
MgO(c)/3 ML Fe/Ag& A & 0.15 & 35.6 & 1.9(1) &0.10(8) & 37.3(9)\\
                 &  B & 0.32 & 31.0 & 5.6(2) &0.43(8) & 57.3(9)\\
                 &  C & 0.32 &  &  & & 5.4(9)\\
MgO(p)/5 ML Fe/Ag& A & 0.16 & 36.0 & 1.6(1) &0.19(8) & 56.1(9)\\
                 &  B & 0.33 & 31.1 & 4.1(2) &0.21(8) & 40.8(9)\\
                 &  C & 0.32 &  &  & & 3.1(9)\\
MgO(c)/5 ML Fe/Ag& A & 0.15 & 36.1 & 1.7(1) &0.13(8) & 43.7(9)\\
                 &  B & 0.32 & 31.0 & 4.8(2) &0.31(8) & 52.9(9)\\
                 &  C & 0.33 &  &  & & 3.4(9)\\
MgO(p)/6 ML Fe/Ag& A & 0.16 & 35.1 & 1.4(1) &3.17(8) & 82.8(9)\\
                 &  B & 0.33 & 30.9 & 3.4(2) &3.48(8) & 15.4(9)\\
                 &  C & 0.34 &  &  & & 1.8(9)\\
MgO(c)/6 ML Fe/Ag&A & 0.16 & 35.6 & 1.5(1) &0.11(8) & 80.3(9)\\
                 &  B & 0.33(1) & 30.8 & 3.6(2) &0.23(8) & 16.9(9)\\
                 &  C & 0.34(2) &  &  & & 2.8(9)\\
\hline
\end{tabular*}
\end{table*}

Two components appear. One can be attributed to pure iron clusters the other to iron at the Fe/Ag interface. At low Fe thickness  the relative intensity of RI$_{Fe}$/RI$_{Fe/Ag}$ is smaller than at thicker Fe layers because  the specific surface area is larger.

Fe grows on MgO(100) epitaxially at room
temperature rotated by 45$^{0}$ in plane. Since the surface free energy of Fe is considerably larger than for MgO, 3D Fe islands  form on MgO surface (Vomer-Weber mechanism). The Fe atoms are sitting above the O$^{2-}$ sites on the MgO(100) surface. \cite{Boubeta1,Meyerheim2}
First, round shaped islands of different sizes are
formed at room temperature deposition \cite{Stroscio,Lairson,Thurmer,Lawler,Jordan,Fahsold,Fahsold2,Reitinger}.
The epitaxially grown Fe films should have a small contraction of the
vertical lattice constant to compensate for the in plane expansion
imposed by the 5.8 percent lattice mismatch between MgO and Fe. Nevertheless, the growth of Fe under slight compression was observed\cite{Li2} The films
become continuous and fully cover the substrate surface at
an 6 ML thickness from morphological and magnetic point of view.
\cite{Boubeta2} Cluster formation of Fe on MgO may also expected because of the diffusion mediated adatom capture and the effect of the Ehrlich Sw\"{o}bel factor.  As it was referred in Introduction section, several authors reported Fe-oxid formation and/or significant interaction between the Fe and the O$
^{2-}$ ion on the MgO/Fe interface. In the later case the s
electron density should decrease at the $^{57}$Fe nucleus because
of the correlation effect of 4s electrons at the Fermi surface and the
oxygen ion modifying the charge distribution at the interface and an increase of isomer shift
should appear.  The M\"{o}ssbauer spectra of the samples with 4, 6 and 8 ML $^{57}$Fe
deposited on the MgO surfaces and  covered with MgO measured at room temperature have
IS values very close to 0 mm/s (Table I.) indicating the same s electron
density as in the Fe-Fe atomic neighborhood. This result exclude
the change of electron density on Fe indicating no any significant interaction between the Fe and the O$^{2-}$ ion at the Fe/MgO interfaces or the formation of Fe-oxide
molecular component. For Ag covered samples, the iron atom interact with Ag atom in the interface. This case, components with  increased $\delta$ values appear (Table II.).  The spectra of Fe deposited on polished and cleaved substrates are different (Fig. 2). On polished substrate magnetic relaxation effect appear indicating smaller islands forming on the surface. This case, more structural defects are at  polished surface than in the cleaved one promoting the formation of more islands. The size of these islands become smaller  than  on cleaved surface. Therefore the blocking temperature for the smaller islands are lower than for larger ones (on cleaved surface). The stabilization processes of islands may result in some disorder including dislocations
\cite{Stin,Stau}. The films are not perfectly ordered in the bcc structure and this "disorder" may change the hyperfine magnetic interaction. Therefore, the magnetic fields values appear with distribution  as it was observed earlier for Ag/Fe/Ag layers\cite{Dezsi}. The average internal magnetic field values for 4 and 6 ML thick samples are larger than for bulk iron indicating enhanced field.
The spectra show different relative line intensities depending on Fe thickness. As it is described in previous caption, at low thickness values, up to 6 Ml Fe the spins are parallel to the lattice normal. At higher thickness values the spins turn to parallel to the surface layer. This behavior is very similar as they were observed for Ag/Fe/Ag layers.
\newline The magnetic exchange coupling in ultra thin layers are different relative to the bulk phases, consequently, the magnitude and orientation of magnetic anisotropies are also different. The magnetic anisotropy energy depends on the layer thickness and on the  temperature. The main sources of magnetic anisotropy are the magnetic dipolar interaction and the spin orbit interaction. For very thin layers the dipolar field interaction is not the larger contribution because the layers mostly form relatively small islands, they can not be considered as a magnetic continuum, the long range interactions do not dominate. Instead, the spin orbit interaction is the dominant factor. Depending on the thickness of the deposited layers in the range of 1-10 ML-s calculations\cite{Draa} really showed the magnitude  of the dipolar contribution is of minor importance, the spin-orbit coupling appears dominant. The thickness dependence of spin orientation  for lower thickness values is parallel with the film normal. The linear variation of the spin direction depending on the temperature for 8 ML thick sample (Fig. 4) indicates the effect of the  linear temperature dependent thermal expansion of the lattice according to result of the analysis of the  temperature dependent magneto-elastic anisotropy.\cite{Aktas} At larger thickness (10 ML), in plane magnetization appears, the spin-orbit coupling becomes weak and quenched.

\section{Conclusions}
M\"{o}ssbauer spectroscopy of $^{57}$Fe provided results on the  atomic
interactions and magnetic properties of Fe layers on MgO(100)/Fe/MgO and on MgO(100)/Fe/Ag.
The results showed that depositing iron on MgO (100)
at room temperature, the Fe and O$^{2-}$ ion interaction is extremely weak and no any FeO form at the Fe/MgO interfaces. Up till now, this is the first system where no electronic density changes take place at the interface between  the transition metal layer deposited on a substrate.  Also, no electron density changes were  observed when MgO was deposited on the surface of already deposited metallic layers. Therefore, MgO cover layer can  save the deposited metallic layer for studies  performed in extra vacuum.  The results  indicate that MgO is an ideal insulator layer as the resistor  between the magnetic iron layers in  tunneling magnetic resistance systems. The studies showed enhanced hyperfine magnetic field in the iron ultra-thin layers with  spin orientations depending on their thickness indicating the decrease of the electronic orbital moment for  thicker iron layers. The higher parameter distribution values of the polished substrates comparing to the cleaved ones proves  the difference in the growth process of the iron on the substrate surface.

\textbf{Acknowledgement}\newline
This work was supported by the Hungarian National Research Fund (OTKA)
project No. K62272.

\end{document}